\documentclass[%
 reprint,
superscriptaddress,
 amsmath,amssymb,
 aps,
]{revtex4-2}

\usepackage{graphicx}
\usepackage{dcolumn}
\usepackage{bm}

\begin{document}

\preprint{APS/123-QED}

\title{Vanadium trimers randomly aligned along the $c$-axis direction in layered LiVO$_2$}

\author{K. Kojima}
\affiliation{Department of Applied Physics, Nagoya University, Nagoya 464-8603, Japan}
\author{N. Katayama}\email{katayama.naoyuki@b.mbox.nagoya-u.ac.jp}
\affiliation{Department of Applied Physics, Nagoya University, Nagoya 464-8603, Japan}
\author{S. Tamura}
\affiliation{Department of Applied Physics, Nagoya University, Nagoya 464-8603, Japan}
\author{M. Shiomi}
\affiliation{Department of Applied Physics, Nagoya University, Nagoya 464-8603, Japan}
\author{H. Sawa}					
\affiliation{Department of Applied Physics, Nagoya University, Nagoya 464-8603, Japan}
\date{\today}

\begin{abstract}
Herein, we discuss the identification of vanadium trimers in layered LiVO$_2$ and its sulfide analog of LiVS$_2$ with two-dimensional triangular lattices. Our comprehensive structural studies using synchrotron X-ray diffraction experiments clarified that vanadium trimers are randomly aligned along the $c$-axis direction in LiVO$_2$, while the long-range ordering of vanadium trimers along the $c$-axis direction appears in LiVS$_2$. Our results solve the longstanding issue of cluster patterns in LiVO$_2$ and provide an experimental basis for identifying the mechanism of trimer formation.

\end{abstract}

\maketitle

The entanglement of multiple electronic degrees of freedom in transition metal compounds frequently results in the spontaneous formation of clusters, called orbital molecules \cite{CuIr2S4,MgTi2O4,AlV2O4-2, Fe3O4,LiVO2_discovery,AlV2O4,LiRh2O4,LiVS2,Li033VS2,Attfield}. Recent structural studies have successfully clarified the complex arrangement of clusters, such as ``octamers" in CuIr$_2$S$_4$ \cite{CuIr2S4}, ``helical dimers" in MgTi$_2$O$_4$ \cite{MgTi2O4}, ``pairs of trimers and tetramers" in AlV$_2$O$_4$ \cite{AlV2O4-2} and magnetic ``trimerons" in Fe$_3$O$_4$ \cite{Fe3O4}. However, in layered LiVO$_2$ with a two dimensional triagular lattice, the pattern of clusters has never been identified despite of the intensive structural studies over a half century \cite{LiVO2_Kobayashi,LiVO2_EXAFS,LiVO2_PDF,LiVO2_single_xray,LiVO2_Tian} since J.B. Goodenough proposed the emergence of ``trimers" based on the discussion of symmetry in 1963 \cite{LiVO2_Goodenough}. Nevertheless, many theoretical efforts have been devoted to clarify the mechanism of ``unidentified" trimer formation \cite{LiVO2_Goodenough, LiVO2_Goodenough-2,LiVO2_Onoda,LiVO2_Pen,LiVO2_Yoshitake,LiVO2_Alaska}, resulting in many possible scenarios including charge-density-wave instability \cite{LiVO2_Goodenough, LiVO2_Goodenough-2}, spin Peierls-type \cite{LiVO2_Onoda}, orbital ordering \cite{LiVO2_Pen} and nesting scenarios under the synergetic effect of Coulomb interactions and trigonal-field splitting \cite{LiVO2_Yoshitake}. Furthermore, the recent studies using pair distribution function (PDF) analysis revealed that the disordered orbital molecules often appear in high temperature paramagnetic phase in such systems, possibly leading to the nonequilibrium physics \cite{Li2RuO3, Attfield}. Therefore, the identification of cluster patterns in LiVO$_2$ has been increasingly required.

Some previous structural studies are not contradictory to the emergence of vanadium trimers in LiVO$_2$. Both the vanadium K-edge X-ray absorption fine structure (EXAFS) and PDF show peaks, which correspond to the neighboring V-V distance, which split into two peaks upon cooling below the transition temperature \cite{LiVO2_EXAFS,LiVO2_PDF}. This is expected to be realized when vanadium trimers appear. Single crystal X-ray and electron diffraction experiments clearly show the emergence of superstructure spots at (1/3,1/3,0) and related positions, which is consistent with the vanadium trimer models \cite{LiVO2_single_xray,LiVO2_Tian}. In contrast to these findings, which indirectly supporting the emergence of trimers in $ab$-plane, little is known about the arrangement of cluster patterns along the $c$-axis direction. Previous single crystal X-ray diffraction studies have claimed that the superstructure peaks are accompanied by diffuse streaks along the $c^*$-axis \cite{LiVO2_single_xray,LiVO2_Onoda-2}, which possibly indicates the absence of long-range ordering of cluster patterns along the $c$-axis direction. We need to consider the origin of diffuse streaks and construct a possible structural model for refinement.

Before addressing this issue, we introduce LiVS$_2$, a sulfide analog of LiVO$_2$, as a reference. While LiVO$_2$ exhibits a paramagnetic insulator to nonmagnetic insulator transition at approximately 490 K \cite{LiVO2_Tian}, LiVS$_2$ exhibits a metal to nonmagnetic insulator transition at approximately 314 K \cite{LiVS2}. In LiVS$_2$, on the basis of electron diffraction and EXAFS experimental results, a trimer pattern similar to LiVO$_2$ is expected to be realized in a low-temperature region \cite{LiVS2}. On contrast, the diffuse scattering along the $c^*$-axis direction has not been identified in LiVS$_2$, expecting us the trimer patterns appear with a long-range ordering along the $c$-axis direction. Therefore, the comprehensive structural studies of both compounds should reveal the underlying factor which generates a long-range ordering of cluster patterns along the $c$-axis direction, leading to a complete understanding of the cluster patterns realized in low temperature phases.

In this letter, we report vanadium trimerization, which commonly appears in low-temperature phases of LiVO$_2$ and LiVS$_2$. While the cluster patterns are long-range along the $c$-axis direction in LiVS$_2$, the long-range ordering is intrinsically absent in LiVO$_2$. We suggest that the parent stacking structures are responsible for the absence/presence of long-range ordering of cluster patterns along the $c$-axis direction and propose a low-temperature structural model for LiVO$_2$, which consists of three possible trimer patterns per VO$_2$ layer. One out of three patterns appears arbitrary in each VO$_2$ layer, resulting the long-range ordering of cluster patterns is intrinsically absent along the $c$-axis direction in LiVO$_2$. The simulated PDF pattern obtained from the structural model fit well the experimentally obtained PDF pattern at 300 K. Our results successfully solve the longstanding issue of the cluster patterns of LiVO$_2$, and provide an experimental basis for identifying the mechanism of trimerization. 

All samples were prepared according to the recipe suggested by Katayama $et$ $al$. \cite{LiVS2} and Tian $et$ $al$. \cite{LiVO2_Tian}. Our synchrotron powder X-ray diffraction experiments clarified that the ratios of Li/V are 1.00(3) for LiVS$_2$ and 0.97(1) LiVO$_2$, respectively. Both samples were confirmed to exhibit clear transitions at the reported temperatures using the synchrotron powder X-ray diffraction experiment. Single crystal X-ray diffraction experiment was performed using R-AXIS RAPID-S (RIGAKU) equipped with Mo target. Synchrotron powder X-ray diffraction experiments with $E$ = 19 keV were performed at BL5S2 beamline equipped at Aichi Synchrotron, Japan. RIETAN-FP and VESTA software were employed for the Rietveld analysis and graphical purpose, respectively \cite{Rietan,VESTA}. PILATUS 100K was used for high-resolution measurement and high-speed data collection. High energy synchrotron X-ray diffraction experiments with $E$ = 61 keV was performed for collecting the data for PDF analysis at BL04B2 at SPring-8, Japan. The hybrid detectors of Ge and CdTe were employed there. The reduced PDF $G$($r$) was obtained by the conventional Fourier transform of the collected data \cite{ohara}. PDFgui package was used for analyzing the $G$($r$) \cite{PDFgui}.


\begin{figure}
\includegraphics[width=80mm]{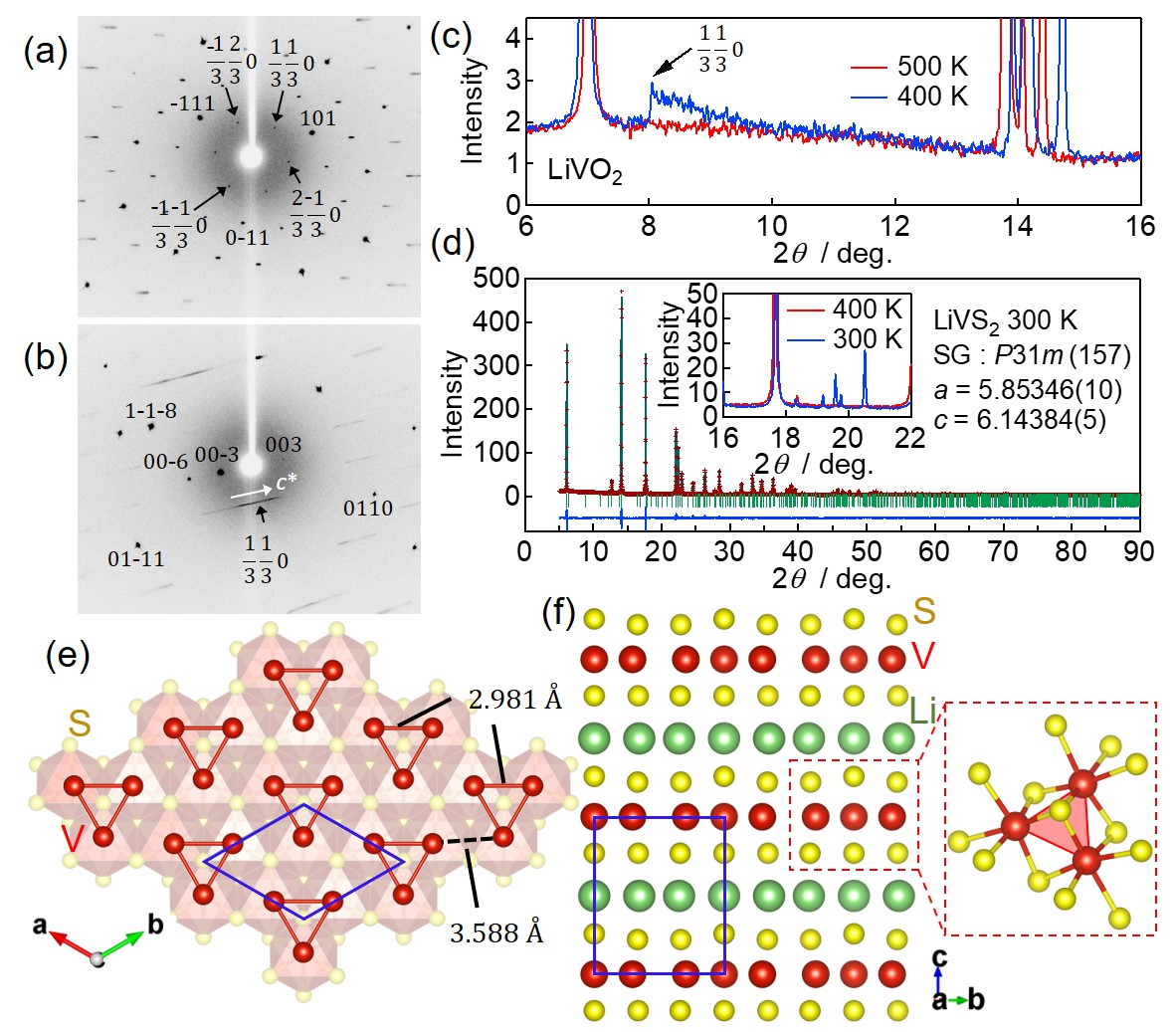}
\caption{\label{fig:Fig1} (a),(b) Single crystal X-ray diffraction patterns of LiVO$_2$ at 300 K for the perpendicular (a) and parallel (b) to $c^*$-direction. (c) Powder diffraction patterns above and below the transition temperature of approximately 490 K in LiVO$_2$. (d) Rietveld refinement of LiVS$_2$ at 300 K, assuming the space group $P$31$m$. The obtained reliability factors were $R_{wp}$ = 5.033 \%, $R_{p}$ = 4.621 \%, $R_{e}$ = 3.181 \% and $S$ = 1.5821. The inset shows powder diffraction patterns above and below the transition temperature of 314 K. (e),(f) Obtained crystal structures of LiVS$_2$ at 300 K. }
\end{figure}

Figures \ref{fig:Fig1}(a) and (b) show the single crystal X-ray diffraction patterns of LiVO$_2$ obtained at 300 K. The superstructure spots appearing at (1/3,1/3,0), and the related positions clearly appear in Figure \ref{fig:Fig1}(a). However, the superstructure spots are accompanied by diffuse streaks without any internal structures along the $c^*$-direction, as shown in \ref{fig:Fig1}(b), while the fundamental peaks remain sharp. As shown in Figure \ref{fig:Fig1}(c), the asymmetric broad superstructure peaks appear in the powder X-ray diffraction data below the transition temperature, consistent with the single crystal X-ray diffraction results. The diffuse streaks appearing accompanied by superstructure spots indicate the absence of long-range ordering of cluster patterns along the $c$-axis direction in LiVO$_2$.

\begin{figure}
\includegraphics[width=80mm]{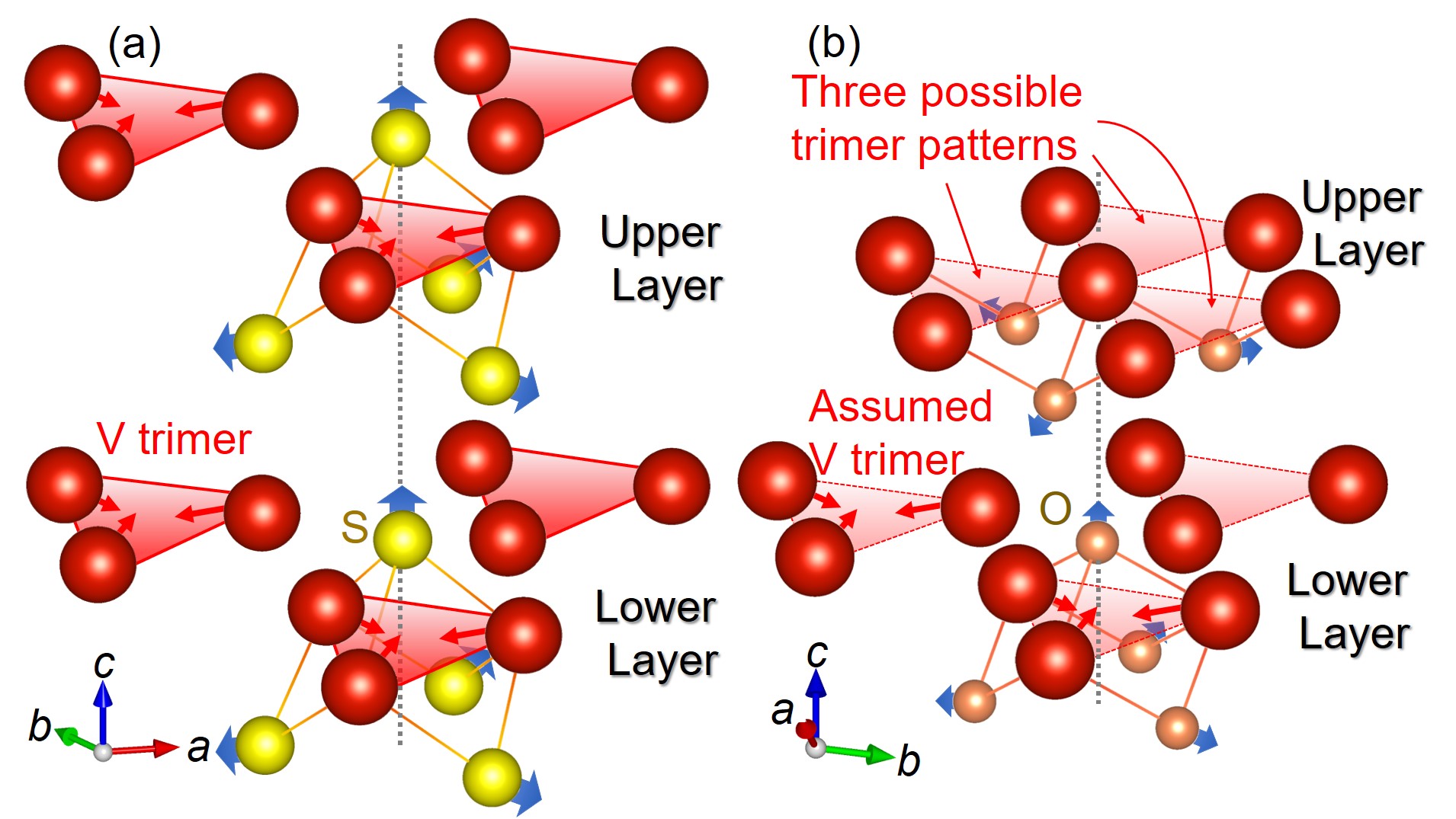}
\caption{\label{fig:Fig2} Schematic pictures of the ($a$) experimentally identified trimer arrangements of LiVS$_2$ and ($b$) expected trimer arrangement of LiVO$_2$.  Li ions are not displayed for simplicity. }
\end{figure}

\begin{figure*}
\includegraphics[width=140mm]{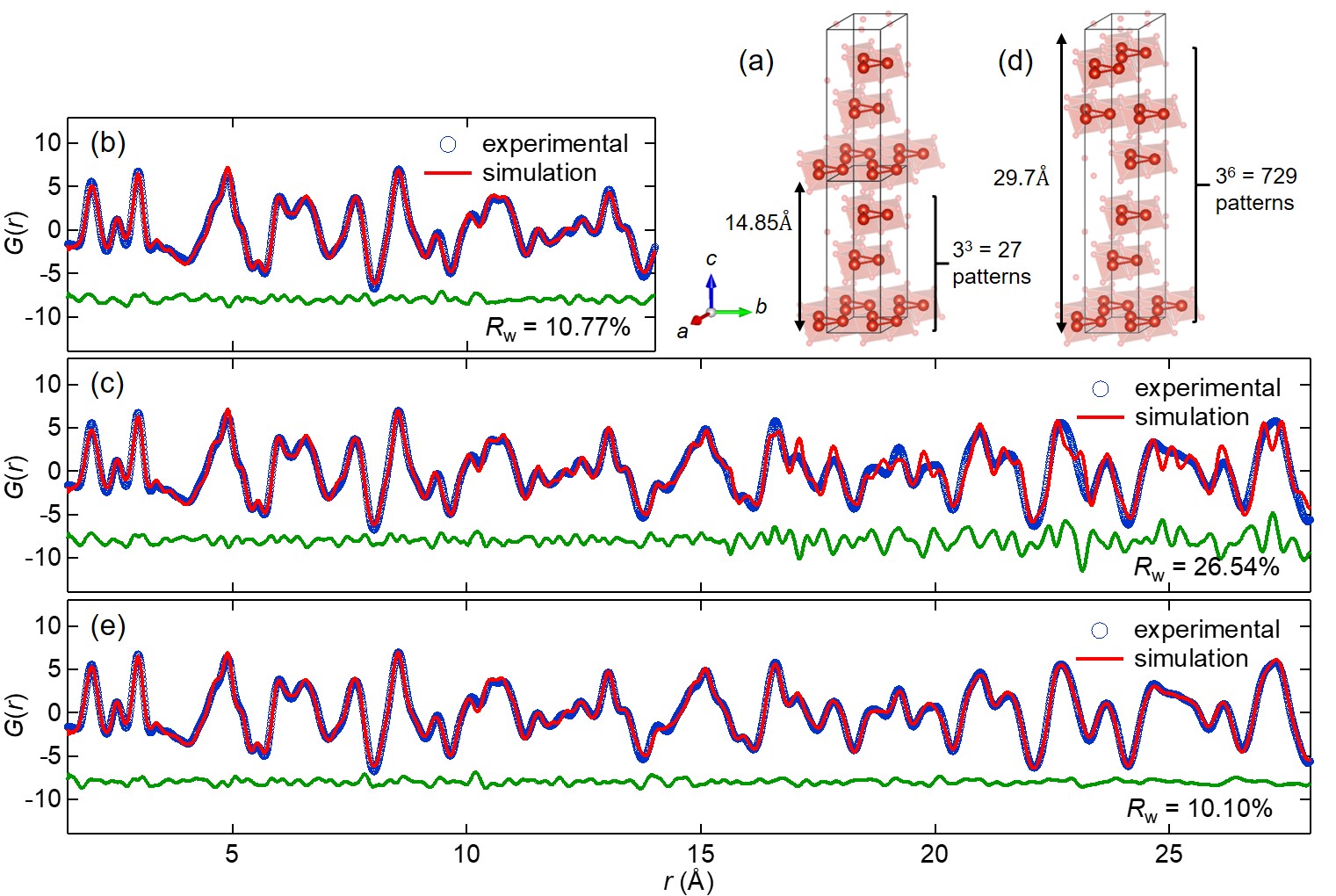}
\caption{\label{fig:Fig3} (a) Schematic picture of one of the 3$^3$ = 27 possible trimer patterns in a unit cell consisting of three VO$_2$ layers. (b),(c) Refined PDF patterns for LiVO$_2$ in the range of 1.8 $\leqq$ $r$(\AA) $\leqq$ 14 (b) and 1.8 $\leqq$ $r$(\AA) $\leqq$ 28 for (c). A simulated pattern obtained from 3$^3$ = 27 possible trimer patterns was employed for the fitting. (d) Schematic picture of one of the 3$^6$ = 729 possible trimer patterns in a unit cell consisting of six VO$_2$ layers. (e) Refined PDF patterns for LiVO$_2$ in the range of 1.8 $\leqq$ $r$(\AA) $\leqq$ 28. A simulated pattern obtained from 3$^6$ = 729 possible trimer patterns was employed for the fitting. }
\end{figure*}

In contrast to LiVO$_2$, the prominent superstructure peaks appear for LiVS$_2$ below the transition temperature, as shown in the inset in Figure \ref{fig:Fig1}(d). This observation indicates the presence of long-range ordering in cluster patterns along the $c$-axis direction in LiVS$_2$. By assuming the trigonal space group $P$31$m$, we can successfully refine the crystal structure to obtain the low temperature crystal structure with vanadium trimers, as shown in Figures \ref{fig:Fig1}(e) and (f). Of note, vanadium trimerization displaces the nearest neighboring sulfur ion upwards due to the increasing Coulomb repulsion between them, which results in an uneven buckling structure of sulfur layers on both sides of the vanadium layer, as shown in the horizontal graph in Figure \ref{fig:Fig1}(f).

It is important to understand what distinguishes LiVO$_2$ from LiVS$_2$ in the absence/presence of long-range ordering of cluster patterns along the $c$-axis direction. Here we explain that the difference in stacking structure among them can be attributed to the absence/presence of long-range ordering. While LiVO$_2$ crystallizes in a 3$c$ structure with $R$$\bar{3}$$m$, LiVS$_2$ possesses a 1$c$ structure with $P$$\bar{3}$$m$1 at high temperatures. When vanadium trimers are formed in Lower Layer as shown in Figure \ref{fig:Fig2}(a), vanadium trimers displace the nearest neighboring sulfur ions upwards due to the Coulomb repulsion. The displaced sulfur ion further displaces the facing sulfur ions in Upper Layer toward the inner-plane direction due to the increasing Coulomb repulsion. This causes further periodic shading of the Coulomb potential on the vanadium sites in the Upper Layer, which leads to the unique arrangement of vanadium trimers in Upper Layer. Specifically, the long-range ordering of trimer patterns along the $c$-axis direction is caused by the propagation of atomic displacement accompanied by change in the Coulomb potential in LiVS$_2$ with a 1$c$ structure. When the trimers are assumed to be realized in Lower Layer of LiVO$_2$ with a 3$c$ stacking structure as shown in Figure \ref{fig:Fig2}(b), the trimers should displace the surrounding oxygen ions in a similar manner to LiVS$_2$. In contrast to LiVS$_2$, however, the trimer configuration in Upper Layer cannot be uniformly determined in LiVO$_2$ because three possible trimer configurations become energetically comparable due to the inherent 3$c$ structure. We expect that the possible three trimer patterns, shown in Figure \ref{fig:Fig2}(b), should cause a kind of frustration, which lead to the intrinsic absence of long-range ordering along the $c$-axis direction in the trimer arrangements of LiVO$_2$. Additional information is supplied in the Supplemental Information.

Based on the abovementioned discussion, we construct the possible structural model on the basis of the following three policies. First, long-ranged vanadium trimer arrangements appear in the $ab$-plane. Second, oxygen ions are displaced in a similar manner to sulfur ions in LiVS$_2$ accompanied by trimerization. Third, there are three possible trimer arrangements per VO$_2$ layer, as shown in Upper Layer of Figure \ref{fig:Fig2}(b), which results in 3$^3$ = 27 possible patterns in a unit cell consisting of three VO$_2$ layers. One of the possible patterns is schematically shown in Figure \ref{fig:Fig3}(a). Once the initial structural model is constructed, we can check the validity of the constructed model from the PDF analysis. If the abovementioned model is correct, we can refine the reduced PDF $G$($r$) data using the average calculated patterns obtained from 27 possible trimer patterns. The refinement details and obtained parameters are summarized in the Supplemental Information.

Figure \ref{fig:Fig3}(b) shows the PDF analysis result of LiVO$_2$ in the range of 1.8 $\leqq$ $r$(\AA) $\leqq$ 14. The $G$($r$) pattern was obtained from the high-energy synchrotron X-ray diffraction data collected at 300 K. It is observed that the calculated results fit the experimental data well with a small residual. However, when we expand the refined $r$ region up to 1.8 $\leqq$ $r$(\AA) $\leqq$ 28, a large residual appears in $r$(\AA) $\geqq$ $\sim$15 and the reliability factor becomes worse, as shown in Figure \ref{fig:Fig3}(c). This result occurs because the current model covers only the range of $r$(\AA) $\leqq$ $\sim$14.85, which corresponds to the thickness of three-layers. In the current structural model, the trimer pattern appears in a unit cell consisting of three VO$_2$ layers and reappears in the neighboring three VO$_2$ layers due to the translational symmetry of the unit cell. This critically reduces the possible trimer patterns and makes the fitting worse in $r$(\AA) $\geqq$ $\sim$15. On the basis of the abovementioned discussion, we fitted the $G$($r$) pattern in the range of 1.8 $\leqq$ $r$(\AA) $\leqq$ 28, which approximately corresponds to the thickness of six-layers, using the average calculated patterns obtained from 3$^6$ = 729 possible trimer patterns, as schematically shown in Figure \ref{fig:Fig3}(d). The result shows a good fit with a small residual, as shown in Figure \ref{fig:Fig3}(e). The abovementioned results clearly indicate our model of the cluster arrangements in LiVO$_2$ is correct, i.e., vanadium trimers crystallize in the $ab$-plane while they are in a glassy state along the $c$-axis direction.

Thus, our structural studies clearly solved the longstanding issue of the trimerization in LiVO$_2$ and show that vanadium trimers are indeed realized both in LiVO$_2$ and LiVS$_2$. The absence of long-range ordering of trimer patterns along the $c$-axis direction in LiVO$_2$ indicates the transition intrinsically appears as a two-dimensional nature, consistent with the fact that cluster formation appears as a clear first-order transition despite the appearance of strong diffuse scattering. Although the arrangement of vanadium trimers along the $c$-axis direction is sharply different between LiVO$_2$ and LiVS$_2$, the ratio of V-V distances between inner- and intra- trimers reaches similar values: 0.848 for LiVO$_2$ at 300 K and 0.831 for LiVS$_2$ at 200 K. This indicates that the strong bonding nature commonly appears both in LiVO$_2$ and LiVS$_2$. Although the mechanism of trimer formation is beyond the scope of this paper, our results provide an experimental basis for the vigorous discussion, and the obtained parameters contribute to further theoretical and experimental understanding. 

\begin{acknowledgments}
The authors are grateful to E. Ishii and K. Ohara for the experimental supports. The work leading to these results has received funding from the Grant in Aid for Scientiﬁc Research (No. JP17K17793). The Thermal and Electric Energy Technology Inc. Foundation, and Daiko Foundation. This work was carried out under the Visiting Researcher’s Program of the Institute for Solid State Physics, the University of Tokyo, and the Collaborative Research Projects of Laboratory for Materials and Structures, Institute of Innovative Research, Tokyo Institute of Technology. The synchrotron powder X-ray diffraction experiments for Rietveld analysis were conducted at the BL5S2 of Aichi Synchrotron Radiation Center, Aichi Science and Technology Foundation, Aichi, Japan (Proposals No. 201704099, No. 201806026, No. 201804016, No. 201901018, No. 201902056). The high energy synchrotron powder X-ray diffraction experiments for PDF analysis were conducted at the BL04B2 of SPring-8, Hyogo, Japan (Proposals No. 2018B1128 and No. 2018B1145). 
\end{acknowledgments}

\appendix

\nocite{*}

\bibliography{LiVO2_trimer_references}

\providecommand{\noopsort}[1]{}\providecommand{\singleletter}[1]{#1}%
\begin{thebibliography}{27}%
\makeatletter
\providecommand \@ifxundefined [1]{%
 \@ifx{#1\undefined}
}%
\providecommand \@ifnum [1]{%
 \ifnum #1\expandafter \@firstoftwo
 \else \expandafter \@secondoftwo
 \fi
}%
\providecommand \@ifx [1]{%
 \ifx #1\expandafter \@firstoftwo
 \else \expandafter \@secondoftwo
 \fi
}%
\providecommand \natexlab [1]{#1}%
\providecommand \enquote  [1]{``#1''}%
\providecommand \bibnamefont  [1]{#1}%
\providecommand \bibfnamefont [1]{#1}%
\providecommand \citenamefont [1]{#1}%
\providecommand \href@noop [0]{\@secondoftwo}%
\providecommand \href [0]{\begingroup \@sanitize@url \@href}%
\providecommand \@href[1]{\@@startlink{#1}\@@href}%
\providecommand \@@href[1]{\endgroup#1\@@endlink}%
\providecommand \@sanitize@url [0]{\catcode `\\12\catcode `\$12\catcode
  `\&12\catcode `\#12\catcode `\^12\catcode `\_12\catcode `\%12\relax}%
\providecommand \@@startlink[1]{}%
\providecommand \@@endlink[0]{}%
\providecommand \url  [0]{\begingroup\@sanitize@url \@url }%
\providecommand \@url [1]{\endgroup\@href {#1}{\urlprefix }}%
\providecommand \urlprefix  [0]{URL }%
\providecommand \Eprint [0]{\href }%
\providecommand \doibase [0]{https://doi.org/}%
\providecommand \selectlanguage [0]{\@gobble}%
\providecommand \bibinfo  [0]{\@secondoftwo}%
\providecommand \bibfield  [0]{\@secondoftwo}%
\providecommand \translation [1]{[#1]}%
\providecommand \BibitemOpen [0]{}%
\providecommand \bibitemStop [0]{}%
\providecommand \bibitemNoStop [0]{.\EOS\space}%
\providecommand \EOS [0]{\spacefactor3000\relax}%
\providecommand \BibitemShut  [1]{\csname bibitem#1\endcsname}%
\let\auto@bib@innerbib\@empty
\bibitem [{\citenamefont {Radaelli}\ \emph {et~al.}(2002)\citenamefont
  {Radaelli}, \citenamefont {Horibe}, \citenamefont {Gutmann}, \citenamefont
  {Ishibashi}, \citenamefont {Chen}, \citenamefont {Ibberson}, \citenamefont
  {Koyama}, \citenamefont {Hor}, \citenamefont {Kiryukhin},\ and\ \citenamefont
  {Cheong}}]{CuIr2S4}%
  \BibitemOpen
  \bibfield  {author} {\bibinfo {author} {\bibfnamefont {P.~G.}\ \bibnamefont
  {Radaelli}}, \bibinfo {author} {\bibfnamefont {Y.}~\bibnamefont {Horibe}},
  \bibinfo {author} {\bibfnamefont {M.~J.}\ \bibnamefont {Gutmann}}, \bibinfo
  {author} {\bibfnamefont {H.}~\bibnamefont {Ishibashi}}, \bibinfo {author}
  {\bibfnamefont {C.~H.}\ \bibnamefont {Chen}}, \bibinfo {author}
  {\bibfnamefont {R.~M.}\ \bibnamefont {Ibberson}}, \bibinfo {author}
  {\bibfnamefont {Y.}~\bibnamefont {Koyama}}, \bibinfo {author} {\bibfnamefont
  {Y.~S.}\ \bibnamefont {Hor}}, \bibinfo {author} {\bibfnamefont
  {V.}~\bibnamefont {Kiryukhin}},\ and\ \bibinfo {author} {\bibfnamefont
  {S.~W.}\ \bibnamefont {Cheong}},\ }\href@noop {} {\bibfield  {journal}
  {\bibinfo  {journal} {Nature (London)}\ }\textbf {\bibinfo {volume} {416}},\
  \bibinfo {pages} {155} (\bibinfo {year} {2002})}\BibitemShut {NoStop}%
\bibitem [{\citenamefont {Schmidt}\ \emph {et~al.}(2004)\citenamefont
  {Schmidt}, \citenamefont {W.~Ratcliff}, \citenamefont {Radaelli},
  \citenamefont {Refson}, \citenamefont {Harrison},\ and\ \citenamefont
  {Cheong}}]{MgTi2O4}%
  \BibitemOpen
  \bibfield  {author} {\bibinfo {author} {\bibfnamefont {M.}~\bibnamefont
  {Schmidt}}, \bibinfo {author} {\bibfnamefont {I.}~\bibnamefont
  {W.~Ratcliff}}, \bibinfo {author} {\bibfnamefont {P.~G.}\ \bibnamefont
  {Radaelli}}, \bibinfo {author} {\bibfnamefont {K.}~\bibnamefont {Refson}},
  \bibinfo {author} {\bibfnamefont {N.~M.}\ \bibnamefont {Harrison}},\ and\
  \bibinfo {author} {\bibfnamefont {S.~W.}\ \bibnamefont {Cheong}},\
  }\href@noop {} {\bibfield  {journal} {\bibinfo  {journal} {Phys. Rev. Lett.}\
  }\textbf {\bibinfo {volume} {92}},\ \bibinfo {pages} {056402} (\bibinfo
  {year} {2004})}\BibitemShut {NoStop}%
\bibitem [{\citenamefont {Browne}\ \emph {et~al.}(2017)\citenamefont {Browne},
  \citenamefont {Kimber},\ and\ \citenamefont {Attﬁeld}}]{AlV2O4-2}%
  \BibitemOpen
  \bibfield  {author} {\bibinfo {author} {\bibfnamefont {A.~J.}\ \bibnamefont
  {Browne}}, \bibinfo {author} {\bibfnamefont {S.~A.~J.}\ \bibnamefont
  {Kimber}},\ and\ \bibinfo {author} {\bibfnamefont {J.~P.}\ \bibnamefont
  {Attﬁeld}},\ }\href@noop {} {\bibfield  {journal} {\bibinfo  {journal}
  {Phys. Rev. Mater.}\ }\textbf {\bibinfo {volume} {1}},\ \bibinfo {pages}
  {052003(R)} (\bibinfo {year} {2017})}\BibitemShut {NoStop}%
\bibitem [{\citenamefont {Senn}\ \emph {et~al.}(2012)\citenamefont {Senn},
  \citenamefont {Wright},\ and\ \citenamefont {Attfield}}]{Fe3O4}%
  \BibitemOpen
  \bibfield  {author} {\bibinfo {author} {\bibfnamefont {M.~S.}\ \bibnamefont
  {Senn}}, \bibinfo {author} {\bibfnamefont {J.~P.}\ \bibnamefont {Wright}},\
  and\ \bibinfo {author} {\bibfnamefont {J.~P.}\ \bibnamefont {Attfield}},\
  }\href@noop {} {\bibfield  {journal} {\bibinfo  {journal} {Nature (London)}\
  }\textbf {\bibinfo {volume} {481}},\ \bibinfo {pages} {173} (\bibinfo {year}
  {2012})}\BibitemShut {NoStop}%
\bibitem [{\citenamefont {R$\ddot{u}$dorff}\ and\ \citenamefont
  {Becker}(1954)}]{LiVO2_discovery}%
  \BibitemOpen
  \bibfield  {author} {\bibinfo {author} {\bibfnamefont {V.~W.}\ \bibnamefont
  {R$\ddot{u}$dorff}}\ and\ \bibinfo {author} {\bibfnamefont {H.}~\bibnamefont
  {Becker}},\ }\href@noop {} {\bibfield  {journal} {\bibinfo  {journal}
  {Zeitschrift f$\ddot{u}$r Naturforschung B}\ }\textbf {\bibinfo {volume}
  {9}},\ \bibinfo {pages} {614} (\bibinfo {year} {1954})}\BibitemShut {NoStop}%
\bibitem [{\citenamefont {Horibe}\ \emph {et~al.}(2006)\citenamefont {Horibe},
  \citenamefont {Shingu}, \citenamefont {Kurushima}, \citenamefont {Ishibashi},
  \citenamefont {Ikeda}, \citenamefont {Kato}, \citenamefont {Motome},
  \citenamefont {Furukawa}, \citenamefont {Mori},\ and\ \citenamefont
  {Katsufuji}}]{AlV2O4}%
  \BibitemOpen
  \bibfield  {author} {\bibinfo {author} {\bibfnamefont {Y.}~\bibnamefont
  {Horibe}}, \bibinfo {author} {\bibfnamefont {M.}~\bibnamefont {Shingu}},
  \bibinfo {author} {\bibfnamefont {K.}~\bibnamefont {Kurushima}}, \bibinfo
  {author} {\bibfnamefont {H.}~\bibnamefont {Ishibashi}}, \bibinfo {author}
  {\bibfnamefont {N.}~\bibnamefont {Ikeda}}, \bibinfo {author} {\bibfnamefont
  {K.}~\bibnamefont {Kato}}, \bibinfo {author} {\bibfnamefont {Y.}~\bibnamefont
  {Motome}}, \bibinfo {author} {\bibfnamefont {N.}~\bibnamefont {Furukawa}},
  \bibinfo {author} {\bibfnamefont {S.}~\bibnamefont {Mori}},\ and\ \bibinfo
  {author} {\bibfnamefont {T.}~\bibnamefont {Katsufuji}},\ }\href@noop {}
  {\bibfield  {journal} {\bibinfo  {journal} {Phys. Rev. Lett.}\ }\textbf
  {\bibinfo {volume} {96}},\ \bibinfo {pages} {086406} (\bibinfo {year}
  {2006})}\BibitemShut {NoStop}%
\bibitem [{\citenamefont {Okamoto}\ \emph {et~al.}(2008)\citenamefont
  {Okamoto}, \citenamefont {Niitaka}, \citenamefont {Uchida}, \citenamefont
  {Waki}, \citenamefont {Takigawa}, \citenamefont {Nakatsu}, \citenamefont
  {Sekiyama}, \citenamefont {Suga}, \citenamefont {Arita},\ and\ \citenamefont
  {Takagi}}]{LiRh2O4}%
  \BibitemOpen
  \bibfield  {author} {\bibinfo {author} {\bibfnamefont {Y.}~\bibnamefont
  {Okamoto}}, \bibinfo {author} {\bibfnamefont {S.}~\bibnamefont {Niitaka}},
  \bibinfo {author} {\bibfnamefont {M.}~\bibnamefont {Uchida}}, \bibinfo
  {author} {\bibfnamefont {T.}~\bibnamefont {Waki}}, \bibinfo {author}
  {\bibfnamefont {M.}~\bibnamefont {Takigawa}}, \bibinfo {author}
  {\bibfnamefont {Y.}~\bibnamefont {Nakatsu}}, \bibinfo {author} {\bibfnamefont
  {A.}~\bibnamefont {Sekiyama}}, \bibinfo {author} {\bibfnamefont
  {S.}~\bibnamefont {Suga}}, \bibinfo {author} {\bibfnamefont {R.}~\bibnamefont
  {Arita}},\ and\ \bibinfo {author} {\bibfnamefont {H.}~\bibnamefont
  {Takagi}},\ }\href@noop {} {\bibfield  {journal} {\bibinfo  {journal} {Phys.
  Rev. Lett.}\ }\textbf {\bibinfo {volume} {101}},\ \bibinfo {pages} {086404}
  (\bibinfo {year} {2008})}\BibitemShut {NoStop}%
\bibitem [{\citenamefont {Katayama}\ \emph {et~al.}(2009)\citenamefont
  {Katayama}, \citenamefont {Uchida}, \citenamefont {Hashizume}, \citenamefont
  {Niitaka}, \citenamefont {Matsuno}, \citenamefont {Matsumura}, \citenamefont
  {Nishihata}, \citenamefont {Mizuki}, \citenamefont {Takeshita}, \citenamefont
  {Gauzzi}, \citenamefont {Nohara},\ and\ \citenamefont {Takagi}}]{LiVS2}%
  \BibitemOpen
  \bibfield  {author} {\bibinfo {author} {\bibfnamefont {N.}~\bibnamefont
  {Katayama}}, \bibinfo {author} {\bibfnamefont {M.}~\bibnamefont {Uchida}},
  \bibinfo {author} {\bibfnamefont {D.}~\bibnamefont {Hashizume}}, \bibinfo
  {author} {\bibfnamefont {S.}~\bibnamefont {Niitaka}}, \bibinfo {author}
  {\bibfnamefont {J.}~\bibnamefont {Matsuno}}, \bibinfo {author} {\bibfnamefont
  {D.}~\bibnamefont {Matsumura}}, \bibinfo {author} {\bibfnamefont
  {Y.}~\bibnamefont {Nishihata}}, \bibinfo {author} {\bibfnamefont
  {J.}~\bibnamefont {Mizuki}}, \bibinfo {author} {\bibfnamefont
  {N.}~\bibnamefont {Takeshita}}, \bibinfo {author} {\bibfnamefont
  {A.}~\bibnamefont {Gauzzi}}, \bibinfo {author} {\bibfnamefont
  {M.}~\bibnamefont {Nohara}},\ and\ \bibinfo {author} {\bibfnamefont
  {H.}~\bibnamefont {Takagi}},\ }\href@noop {} {\bibfield  {journal} {\bibinfo
  {journal} {Phys. Rev. Lett.}\ }\textbf {\bibinfo {volume} {103}},\ \bibinfo
  {pages} {146405} (\bibinfo {year} {2009})}\BibitemShut {NoStop}%
\bibitem [{\citenamefont {Katayama}\ \emph {et~al.}(2018)\citenamefont
  {Katayama}, \citenamefont {Tamura}, \citenamefont {Yamaguchi}, \citenamefont
  {Sugimoto}, \citenamefont {Iida}, \citenamefont {Matsukawa}, \citenamefont
  {Hoshikawa}, \citenamefont {Ishigaki}, \citenamefont {Kobayashi},
  \citenamefont {Ohta},\ and\ \citenamefont {Sawa}}]{Li033VS2}%
  \BibitemOpen
  \bibfield  {author} {\bibinfo {author} {\bibfnamefont {N.}~\bibnamefont
  {Katayama}}, \bibinfo {author} {\bibfnamefont {S.}~\bibnamefont {Tamura}},
  \bibinfo {author} {\bibfnamefont {T.}~\bibnamefont {Yamaguchi}}, \bibinfo
  {author} {\bibfnamefont {K.}~\bibnamefont {Sugimoto}}, \bibinfo {author}
  {\bibfnamefont {K.}~\bibnamefont {Iida}}, \bibinfo {author} {\bibfnamefont
  {T.}~\bibnamefont {Matsukawa}}, \bibinfo {author} {\bibfnamefont
  {A.}~\bibnamefont {Hoshikawa}}, \bibinfo {author} {\bibfnamefont
  {T.}~\bibnamefont {Ishigaki}}, \bibinfo {author} {\bibfnamefont
  {S.}~\bibnamefont {Kobayashi}}, \bibinfo {author} {\bibfnamefont
  {Y.}~\bibnamefont {Ohta}},\ and\ \bibinfo {author} {\bibfnamefont
  {H.}~\bibnamefont {Sawa}},\ }\href@noop {} {\bibfield  {journal} {\bibinfo
  {journal} {Phys. Rev. B}\ }\textbf {\bibinfo {volume} {98}},\ \bibinfo
  {pages} {081104(R)} (\bibinfo {year} {2018})}\BibitemShut {NoStop}%
\bibitem [{\citenamefont {Attfield}(2015)}]{Attfield}%
  \BibitemOpen
  \bibfield  {author} {\bibinfo {author} {\bibfnamefont {J.~P.}\ \bibnamefont
  {Attfield}},\ }\href@noop {} {\bibfield  {journal} {\bibinfo  {journal} {APL
  Mater.}\ }\textbf {\bibinfo {volume} {3}},\ \bibinfo {pages} {041510}
  (\bibinfo {year} {2015})}\BibitemShut {NoStop}%
\bibitem [{\citenamefont {Kobayashi}\ \emph {et~al.}(1969)\citenamefont
  {Kobayashi}, \citenamefont {Kosuge},\ and\ \citenamefont
  {Kachi}}]{LiVO2_Kobayashi}%
  \BibitemOpen
  \bibfield  {author} {\bibinfo {author} {\bibfnamefont {K.}~\bibnamefont
  {Kobayashi}}, \bibinfo {author} {\bibfnamefont {K.}~\bibnamefont {Kosuge}},\
  and\ \bibinfo {author} {\bibfnamefont {S.}~\bibnamefont {Kachi}},\
  }\href@noop {} {\bibfield  {journal} {\bibinfo  {journal} {Mater. Res.
  Bull.}\ }\textbf {\bibinfo {volume} {4}},\ \bibinfo {pages} {95} (\bibinfo
  {year} {1969})}\BibitemShut {NoStop}%
\bibitem [{\citenamefont {Imai}\ \emph {et~al.}(1995)\citenamefont {Imai},
  \citenamefont {Koike}, \citenamefont {Sawa},\ and\ \citenamefont
  {Takei}}]{LiVO2_EXAFS}%
  \BibitemOpen
  \bibfield  {author} {\bibinfo {author} {\bibfnamefont {K.}~\bibnamefont
  {Imai}}, \bibinfo {author} {\bibfnamefont {M.}~\bibnamefont {Koike}},
  \bibinfo {author} {\bibfnamefont {H.}~\bibnamefont {Sawa}},\ and\ \bibinfo
  {author} {\bibfnamefont {H.}~\bibnamefont {Takei}},\ }\href@noop {}
  {\bibfield  {journal} {\bibinfo  {journal} {J. Solid State Chem.}\ }\textbf
  {\bibinfo {volume} {114}},\ \bibinfo {pages} {184} (\bibinfo {year}
  {1995})}\BibitemShut {NoStop}%
\bibitem [{\citenamefont {Pourpoint}\ \emph {et~al.}(2012)\citenamefont
  {Pourpoint}, \citenamefont {Hua}, \citenamefont {Middlemiss}, \citenamefont
  {Adamson}, \citenamefont {Wang}, \citenamefont {Bruce},\ and\ \citenamefont
  {Grey}}]{LiVO2_PDF}%
  \BibitemOpen
  \bibfield  {author} {\bibinfo {author} {\bibfnamefont {F.}~\bibnamefont
  {Pourpoint}}, \bibinfo {author} {\bibfnamefont {X.}~\bibnamefont {Hua}},
  \bibinfo {author} {\bibfnamefont {D.~S.}\ \bibnamefont {Middlemiss}},
  \bibinfo {author} {\bibfnamefont {P.}~\bibnamefont {Adamson}}, \bibinfo
  {author} {\bibfnamefont {D.}~\bibnamefont {Wang}}, \bibinfo {author}
  {\bibfnamefont {P.~G.}\ \bibnamefont {Bruce}},\ and\ \bibinfo {author}
  {\bibfnamefont {C.~P.}\ \bibnamefont {Grey}},\ }\href@noop {} {\bibfield
  {journal} {\bibinfo  {journal} {Chem. Mater.}\ }\textbf {\bibinfo {volume}
  {24}},\ \bibinfo {pages} {2880} (\bibinfo {year} {2012})}\BibitemShut
  {NoStop}%
\bibitem [{\citenamefont {Imai}\ \emph {et~al.}(1993)\citenamefont {Imai},
  \citenamefont {Koike}, \citenamefont {Sawa},\ and\ \citenamefont
  {Takei}}]{LiVO2_single_xray}%
  \BibitemOpen
  \bibfield  {author} {\bibinfo {author} {\bibfnamefont {K.}~\bibnamefont
  {Imai}}, \bibinfo {author} {\bibfnamefont {M.}~\bibnamefont {Koike}},
  \bibinfo {author} {\bibfnamefont {H.}~\bibnamefont {Sawa}},\ and\ \bibinfo
  {author} {\bibfnamefont {H.}~\bibnamefont {Takei}},\ }\href@noop {}
  {\bibfield  {journal} {\bibinfo  {journal} {J. Solid State Chem.}\ }\textbf
  {\bibinfo {volume} {102}},\ \bibinfo {pages} {277} (\bibinfo {year}
  {1993})}\BibitemShut {NoStop}%
\bibitem [{\citenamefont {Tian}\ \emph {et~al.}(2004)\citenamefont {Tian},
  \citenamefont {Chisholm}, \citenamefont {Khalifah}, \citenamefont {Jin},
  \citenamefont {Sales}, \citenamefont {Nagler},\ and\ \citenamefont
  {Mandrus}}]{LiVO2_Tian}%
  \BibitemOpen
  \bibfield  {author} {\bibinfo {author} {\bibfnamefont {W.}~\bibnamefont
  {Tian}}, \bibinfo {author} {\bibfnamefont {M.~F.}\ \bibnamefont {Chisholm}},
  \bibinfo {author} {\bibfnamefont {P.~G.}\ \bibnamefont {Khalifah}}, \bibinfo
  {author} {\bibfnamefont {R.}~\bibnamefont {Jin}}, \bibinfo {author}
  {\bibfnamefont {B.~C.}\ \bibnamefont {Sales}}, \bibinfo {author}
  {\bibfnamefont {S.~E.}\ \bibnamefont {Nagler}},\ and\ \bibinfo {author}
  {\bibfnamefont {D.}~\bibnamefont {Mandrus}},\ }\href@noop {} {\bibfield
  {journal} {\bibinfo  {journal} {Mater. Res. Bull.}\ }\textbf {\bibinfo
  {volume} {39}},\ \bibinfo {pages} {1319} (\bibinfo {year}
  {2004})}\BibitemShut {NoStop}%
\bibitem [{\citenamefont {Goodenough}(1963)}]{LiVO2_Goodenough}%
  \BibitemOpen
  \bibfield  {author} {\bibinfo {author} {\bibfnamefont {J.~B.}\ \bibnamefont
  {Goodenough}},\ }\href@noop {} {\emph {\bibinfo {title} {Magnetism and the
  Chemical Bond}}}\ (\bibinfo  {publisher} {Interscience and John Wiley,
  N.Y.},\ \bibinfo {year} {1963})\BibitemShut {NoStop}%
\bibitem [{\citenamefont {Goodenough}\ \emph {et~al.}(1991)\citenamefont
  {Goodenough}, \citenamefont {Dutta},\ and\ \citenamefont
  {Manthiram}}]{LiVO2_Goodenough-2}%
  \BibitemOpen
  \bibfield  {author} {\bibinfo {author} {\bibfnamefont {J.~B.}\ \bibnamefont
  {Goodenough}}, \bibinfo {author} {\bibfnamefont {G.}~\bibnamefont {Dutta}},\
  and\ \bibinfo {author} {\bibfnamefont {A.}~\bibnamefont {Manthiram}},\
  }\href@noop {} {\bibfield  {journal} {\bibinfo  {journal} {Phys. Rev. B}\
  }\textbf {\bibinfo {volume} {43}},\ \bibinfo {pages} {10170} (\bibinfo {year}
  {1991})}\BibitemShut {NoStop}%
\bibitem [{\citenamefont {Onoda}\ \emph {et~al.}(1991)\citenamefont {Onoda},
  \citenamefont {Naka},\ and\ \citenamefont {Nagasawa}}]{LiVO2_Onoda}%
  \BibitemOpen
  \bibfield  {author} {\bibinfo {author} {\bibfnamefont {M.}~\bibnamefont
  {Onoda}}, \bibinfo {author} {\bibfnamefont {T.}~\bibnamefont {Naka}},\ and\
  \bibinfo {author} {\bibfnamefont {H.}~\bibnamefont {Nagasawa}},\ }\href@noop
  {} {\bibfield  {journal} {\bibinfo  {journal} {J. Phys. Soc. Jpn.}\ }\textbf
  {\bibinfo {volume} {60}},\ \bibinfo {pages} {2550} (\bibinfo {year}
  {1991})}\BibitemShut {NoStop}%
\bibitem [{\citenamefont {Pen}\ \emph {et~al.}(1997)\citenamefont {Pen},
  \citenamefont {van~den Brink}, \citenamefont {Khomskii},\ and\ \citenamefont
  {Sawatzky}}]{LiVO2_Pen}%
  \BibitemOpen
  \bibfield  {author} {\bibinfo {author} {\bibfnamefont {H.~F.}\ \bibnamefont
  {Pen}}, \bibinfo {author} {\bibfnamefont {J.}~\bibnamefont {van~den Brink}},
  \bibinfo {author} {\bibfnamefont {D.~I.}\ \bibnamefont {Khomskii}},\ and\
  \bibinfo {author} {\bibfnamefont {G.~A.}\ \bibnamefont {Sawatzky}},\
  }\href@noop {} {\bibfield  {journal} {\bibinfo  {journal} {Phys. Rev. Lett.}\
  }\textbf {\bibinfo {volume} {78}},\ \bibinfo {pages} {1323} (\bibinfo {year}
  {1997})}\BibitemShut {NoStop}%
\bibitem [{\citenamefont {Yoshitake}\ and\ \citenamefont
  {Motome}(2011)}]{LiVO2_Yoshitake}%
  \BibitemOpen
  \bibfield  {author} {\bibinfo {author} {\bibfnamefont {J.}~\bibnamefont
  {Yoshitake}}\ and\ \bibinfo {author} {\bibfnamefont {Y.}~\bibnamefont
  {Motome}},\ }\href@noop {} {\bibfield  {journal} {\bibinfo  {journal} {J.
  Phys. Soc. Jpn.}\ }\textbf {\bibinfo {volume} {80}},\ \bibinfo {pages}
  {073711} (\bibinfo {year} {2011})}\BibitemShut {NoStop}%
\bibitem [{\citenamefont {Alaska}(2017)}]{LiVO2_Alaska}%
  \BibitemOpen
  \bibfield  {author} {\bibinfo {author} {\bibfnamefont {S.}~\bibnamefont
  {Alaska}},\ }\href@noop {} {\bibfield  {journal} {\bibinfo  {journal} {Phys.
  Rev. B}\ }\textbf {\bibinfo {volume} {95}},\ \bibinfo {pages} {214119}
  (\bibinfo {year} {2017})}\BibitemShut {NoStop}%
\bibitem [{\citenamefont {Kimber}\ \emph {et~al.}(2014)\citenamefont {Kimber},
  \citenamefont {Mazin}, \citenamefont {Shen}, \citenamefont {Jeschke},
  \citenamefont {Streltsov}, \citenamefont {Argyriou}, \citenamefont
  {Valent$\acute{i}$},\ and\ \citenamefont {Khomskii}}]{Li2RuO3}%
  \BibitemOpen
  \bibfield  {author} {\bibinfo {author} {\bibfnamefont {S.~A.~J.}\
  \bibnamefont {Kimber}}, \bibinfo {author} {\bibfnamefont {I.~I.}\
  \bibnamefont {Mazin}}, \bibinfo {author} {\bibfnamefont {J.}~\bibnamefont
  {Shen}}, \bibinfo {author} {\bibfnamefont {H.~O.}\ \bibnamefont {Jeschke}},
  \bibinfo {author} {\bibfnamefont {S.~V.}\ \bibnamefont {Streltsov}}, \bibinfo
  {author} {\bibfnamefont {D.~N.}\ \bibnamefont {Argyriou}}, \bibinfo {author}
  {\bibfnamefont {R.}~\bibnamefont {Valent$\acute{i}$}},\ and\ \bibinfo
  {author} {\bibfnamefont {D.~I.}\ \bibnamefont {Khomskii}},\ }\href@noop {}
  {\bibfield  {journal} {\bibinfo  {journal} {Phys. Rev. B}\ }\textbf {\bibinfo
  {volume} {89}},\ \bibinfo {pages} {081408(R)} (\bibinfo {year}
  {2014})}\BibitemShut {NoStop}%
\bibitem [{\citenamefont {Onoda}\ and\ \citenamefont
  {Inabe}(1993)}]{LiVO2_Onoda-2}%
  \BibitemOpen
  \bibfield  {author} {\bibinfo {author} {\bibfnamefont {M.}~\bibnamefont
  {Onoda}}\ and\ \bibinfo {author} {\bibfnamefont {T.}~\bibnamefont {Inabe}},\
  }\href@noop {} {\bibfield  {journal} {\bibinfo  {journal} {J. Phys. Soc.
  Jpn.}\ }\textbf {\bibinfo {volume} {62}},\ \bibinfo {pages} {2216} (\bibinfo
  {year} {1993})}\BibitemShut {NoStop}%
\bibitem [{\citenamefont {Izumi}\ and\ \citenamefont {Momma}(2007)}]{Rietan}%
  \BibitemOpen
  \bibfield  {author} {\bibinfo {author} {\bibfnamefont {F.}~\bibnamefont
  {Izumi}}\ and\ \bibinfo {author} {\bibfnamefont {K.}~\bibnamefont {Momma}},\
  }\href@noop {} {\bibfield  {journal} {\bibinfo  {journal} {Solid State
  Phenom.}\ }\textbf {\bibinfo {volume} {130}},\ \bibinfo {pages} {15}
  (\bibinfo {year} {2007})}\BibitemShut {NoStop}%
\bibitem [{\citenamefont {Momma}\ and\ \citenamefont {Izumi}(2011)}]{VESTA}%
  \BibitemOpen
  \bibfield  {author} {\bibinfo {author} {\bibfnamefont {K.}~\bibnamefont
  {Momma}}\ and\ \bibinfo {author} {\bibfnamefont {F.}~\bibnamefont {Izumi}},\
  }\href@noop {} {\bibfield  {journal} {\bibinfo  {journal} {J. Appl.
  Crystallogr.}\ }\textbf {\bibinfo {volume} {44}},\ \bibinfo {pages} {1272}
  (\bibinfo {year} {2011})}\BibitemShut {NoStop}%
\bibitem [{\citenamefont {Ohara}\ \emph {et~al.}(2018)\citenamefont {Ohara},
  \citenamefont {Tominaka}, \citenamefont {Yamada}, \citenamefont {Takahashi},
  \citenamefont {Yamaguchi}, \citenamefont {Utsuno}, \citenamefont {Umeki},
  \citenamefont {Yao}, \citenamefont {Nakada},\ and\ \citenamefont
  {Takemoto}}]{ohara}%
  \BibitemOpen
  \bibfield  {author} {\bibinfo {author} {\bibfnamefont {K.}~\bibnamefont
  {Ohara}}, \bibinfo {author} {\bibfnamefont {S.}~\bibnamefont {Tominaka}},
  \bibinfo {author} {\bibfnamefont {H.}~\bibnamefont {Yamada}}, \bibinfo
  {author} {\bibfnamefont {M.}~\bibnamefont {Takahashi}}, \bibinfo {author}
  {\bibfnamefont {H.}~\bibnamefont {Yamaguchi}}, \bibinfo {author}
  {\bibfnamefont {F.}~\bibnamefont {Utsuno}}, \bibinfo {author} {\bibfnamefont
  {T.}~\bibnamefont {Umeki}}, \bibinfo {author} {\bibfnamefont
  {A.}~\bibnamefont {Yao}}, \bibinfo {author} {\bibfnamefont {K.}~\bibnamefont
  {Nakada}},\ and\ \bibinfo {author} {\bibfnamefont {M.}~\bibnamefont
  {Takemoto}},\ }\href@noop {} {\bibfield  {journal} {\bibinfo  {journal} {J.
  Synchrotron Rad.}\ }\textbf {\bibinfo {volume} {25}},\ \bibinfo {pages}
  {1627} (\bibinfo {year} {2018})}\BibitemShut {NoStop}%
\bibitem [{\citenamefont {Farrow}\ \emph {et~al.}(2007)\citenamefont {Farrow},
  \citenamefont {Juhas}, \citenamefont {Liu}, \citenamefont {Bryndin},
  \citenamefont {Bo$\breve{z}$in}, \citenamefont {Bloch}, \citenamefont
  {Proffen},\ and\ \citenamefont {Billing}}]{PDFgui}%
  \BibitemOpen
  \bibfield  {author} {\bibinfo {author} {\bibfnamefont {C.~L.}\ \bibnamefont
  {Farrow}}, \bibinfo {author} {\bibfnamefont {P.}~\bibnamefont {Juhas}},
  \bibinfo {author} {\bibfnamefont {J.~W.}\ \bibnamefont {Liu}}, \bibinfo
  {author} {\bibfnamefont {D.}~\bibnamefont {Bryndin}}, \bibinfo {author}
  {\bibfnamefont {E.~S.}\ \bibnamefont {Bo$\breve{z}$in}}, \bibinfo {author}
  {\bibfnamefont {J.}~\bibnamefont {Bloch}}, \bibinfo {author} {\bibfnamefont
  {T.}~\bibnamefont {Proffen}},\ and\ \bibinfo {author} {\bibfnamefont
  {S.~J.~L.}\ \bibnamefont {Billing}},\ }\href@noop {} {\bibfield  {journal}
  {\bibinfo  {journal} {J. Phys.: Condens. Matter}\ }\textbf {\bibinfo {volume}
  {19}},\ \bibinfo {pages} {335219} (\bibinfo {year} {2007})}\BibitemShut
  {NoStop}%
\end{thebibliography}%

\end{document}


\preprint{APS/123-QED}

\title{Supplemental Information for \\
Vanadium trimers randomly aligned along the $c$-axis direction in layered LiVO$_2$}

\author{K. Kojima}
\affiliation{Department of Applied Physics, Nagoya University, Nagoya 464-8603, Japan}
\author{N. Katayama}\email{katayama.naoyuki@b.mbox.nagoya-u.ac.jp}
\affiliation{Department of Applied Physics, Nagoya University, Nagoya 464-8603, Japan}
\author{S. Tamura}
\affiliation{Department of Applied Physics, Nagoya University, Nagoya 464-8603, Japan}
\author{M. Shiomi}
\affiliation{Department of Applied Physics, Nagoya University, Nagoya 464-8603, Japan}
\author{H. Sawa}					
\affiliation{Department of Applied Physics, Nagoya University, Nagoya 464-8603, Japan}

\begin{abstract}
In this supplemental information, we supply the additional discussion about the possible trimer arrangement in LiVO$_2$ and the detailed conditions to refine the crystal structures of LiVO$_2$ and LiVS$_2$. The corresponding cif files are also available.

\end{abstract}

\date{\today}
\maketitle
\newpage

{\bf Safely Exclusion of three  ``inverse trimer patterns" from the candidates}

\begin{figure}
\renewcommand{\thefigure}{S1}
\includegraphics[width=170mm]{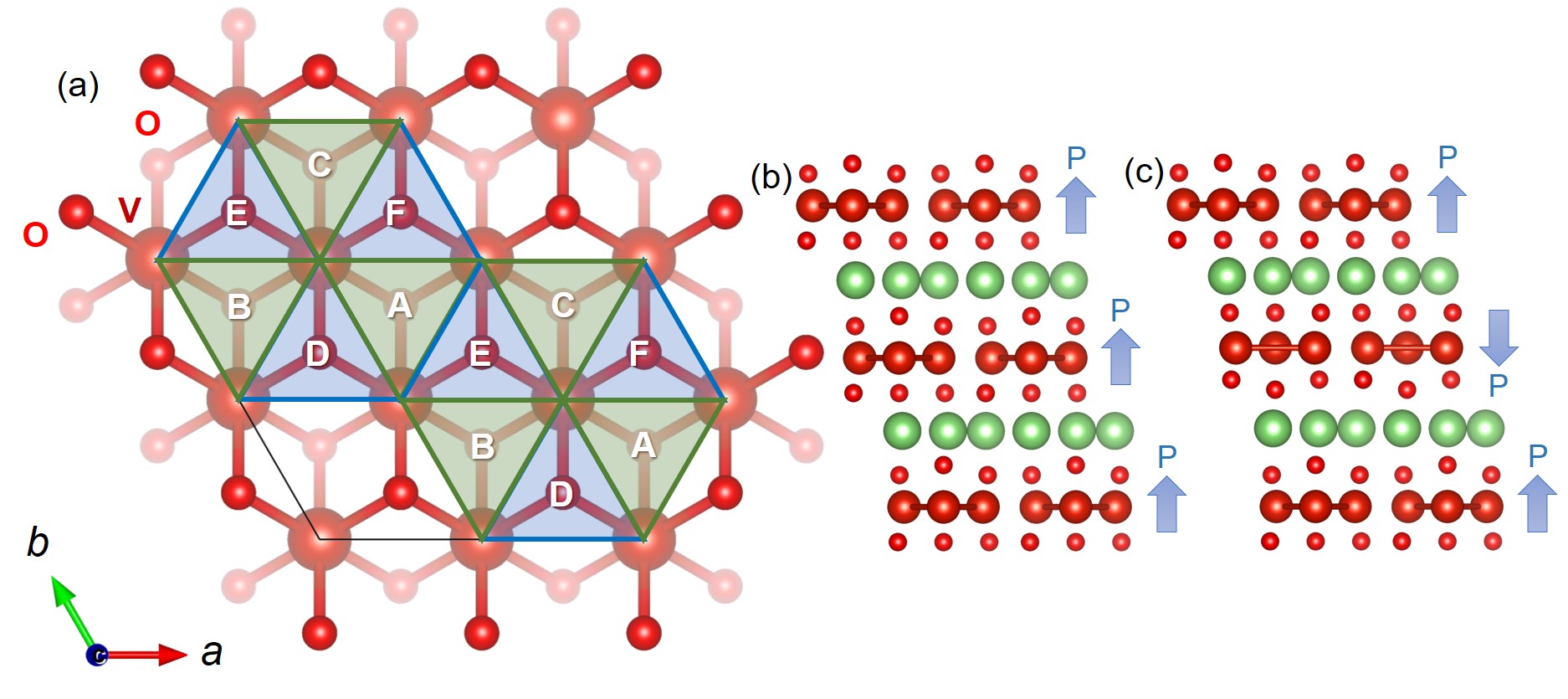}
\caption{\label{fig:FigS1} (a) Possible six trimer patterns consisting of ``regular trimer patterns" (A-C) and ``inverse trimer patterns" (D-F). (b) Ferro-type polarization caused by the regular trimer patterns. (c) Antiparallel polarization caused by the contamination of ``inverse trimer patterns". }
\end{figure}

Herein, we present the additional information about the possible trimer patterns in LiVO$_2$, discussed in the main text. In the main text, we discussed that there are three possible trimer patterns per VO$_2$ layer, and successfully refined the crystal structure using PDF analysis. However, in principle, we can construct six possible trimer patterns in each VO$_2$ layer, as represented as A-F in Figure \ref{fig:FigS1}(a). Here, we discuss that we can safely exclude three out of six patterns based on our single crystalline x-ray diffraction experimental result, shown in Figure 1(b) in the main text.

As schematically shown in Figure \ref{fig:FigS1}(a), there are six possible trimer patterns per VO$_2$ layer. Here, we call the A-C patterns as ``regular trimer patterns" and D-F patterns as ``inverse trimer patterns". In the main text, we considered ``regular trimer patterns" always appear in all VO$_2$ layers. If this happens, the polarization caused by the uneven buckling of oxygen layers on both sides to the vanadium layer appears toward the same direction in all VO$_2$ layers. That is, the ferro-type polarization should appear among VO$_2$ layers, as shown in Figure \ref{fig:FigS1}(b). When it happens, the interaction between neighboring layers retain constant, leading to the constant distance between neighboring VO$_2$ layers. On contrast, if the ``inverse trimer patterns" are included, the ``inverse trimer patterns" definitely causes the antiparallel polarization, as shown in Figure \ref{fig:FigS1}(c) which inevitably generate the inconstant interaction between neighboring VO$_2$ layers. Once it happens, the distance between the neighboring layers becomes inconstant, should generate diffuse scattering and/or superstructure spots in fundamental 00$l$ peaks aligned along the $c^*$-direction. However, we can clearly exclude the latter possibility depending on our single crystalline x-ray diffraction patterns shown in Figure 1(b), where we can clearly find the sharp 00$l$ fundamental peaks without showing any diffuse scattering and/or superstructure peaks. This indicates that the ``inverse trimer patterns" are not realized in LiVO$_2$, therefore, we can safely exclude three D-F trimer patterns from the candidates.



\newpage

{\bf Construction of structural model for PDF refinement}

Here, we present the refinement details employed for PDF analysis of LiVO$_2$. With the reference to the parameters obtained from Rietveld analysis of LiVS$_2$, the restraints were determined to perform the PDF analysis of LiVO$_2$. The obtained parameters are summarized in cif files, which are attached to this letter.

Note that we attained some alart A and B when we apply checkCIF (https://checkcif.iucr.org/) to the cif file of LiVO$_2$. We attach the alert lists with the reasons at the end of this Supplemental Information. There are no significant alerts A and B for the cif file of LiVS$_2$.

\begin{table*}[h]
\renewcommand{\thetable}{SI}
\caption{The table presents the refined atomic coordinates for LiVS$_2$ obtained from the Rielveld analysis of synchroton X-ray powder diffraction data. The cif file is attached in this letter.}
\begin{ruledtabular}
\begin{tabular}{ccccc}
  &Site symmetry&&Coordinate\\ \hline
 &$..m$ $(3c)$&$x$,0,$z$&0,$x$,$z$&$\bar{x}$ $\bar{x}$,$z$\\
 $P31m$&$3..$ $(2b)$&1/3,2/3,$z$&2/3,1/3,$z$\\
 &$3.m$ $(1a)$&0,0,$z$\\ \hline \hline
 & Atoms && Atomic Coordinates\\
 && $x$ & $y$ & $z$\\ \hline
 LiVS$_2$
 &Li ($..m$)&0.34664(86)&0&0.497(15)\\
 &V ($..m$)&0.29448(8)&0&0\\
 & S$_1$ $(..m)$ & 0.66163(18) & 0 & 0.76839(60)\\
 &S$_2$ $(3..)$ &1/3&2/3&0.22019(62)\\
 &S$_3$ $(3.m)$ &0&0&0.25633(59)\\
\end{tabular}
\end{ruledtabular}
\end{table*}

\newpage

\begin{table*}
\renewcommand{\thetable}{SII}
\caption{The table presents the refined atomic coordinates with the restraints for one of the twenty seven possible structures of LiVO$_2$. Restraints were determined with the reference to the refined parameters of LiVS$_2$, presented in Table SI. Note that the unit cell contains three VO$_2$ layers. In the table, the coordinates of one third of composing ions in the unit cell are supplied. For the atomic coordinates of other two third, add (1/3, 1/3, 1/3) and (2/3, 2/3, 2/3) to the parameters supplied in the table. In order to construct other possible twenty six trimer patterns, add (1/3, 2/3, 0) or (2/3, 1/3, 0) to the coordinates of ions composing of each Li and VO$_2$ layers. The combinations should lead to the possible twenty seven structural models in total. The parameters were obtained by PDF analysis of reduced $G$($r$) data obtained from the high energy X-ray diffraction experiment. The result is shown in Figure 3 (b) in the main text with the obtained parameters are x$_{O_1}$ = 0.648609(740), $z_{O_1}$ = 0, $z_{O_2}$ = 0.151166(120), $z_{O_3}$ = 0.162764(120), $x_V$ = 0.298449(49), $z_V$ = 0.077528(130), $x_{Li}$ = 0.351(24) and $z_{Li}$ = 0.2469(42). The cif file is also available.}
\begin{ruledtabular}
\begin{tabular}{ccccc}
 & Atoms && Atomic Coordinates\\
 && $x$ & $y$ & $z$\\ \hline
 LiVO$_2$ (1st Layer)
 & O$_1$ & x$_{O_1}$ & 0 & $z_{O_1}$\\
 & O$_1$ & 0 & $x_{O_1}$ & $z_{O_1}$\\
 & O$_1$ &$\bar{x}_{O_1}$ &$\bar{x}_{O_1}$ & $z_{O_1}$\\
 & V & $x_V$ & 0 & $z_V$\\
 & V & 0 & $x_V$ & $z_V$\\
 & V &$\bar{x}_V$ & $\bar{x}_V$ & $z_V$\\
 & O$_2$ &1/3&2/3&$z_{O_2}$\\
 & O$_2$ &2/3&1/3&$z_{O_2}$\\
 & O$_3$ &0 &0 &$z_{O_3}$\\
 & Li & $x_{Li}+2/3$ & 0+2/3 & $z_{Li}$\\
 & Li & 0+2/3 & $x_{Li}+2/3$ & $z_{Li}$\\
 & Li & $\bar{x}_{Li}+2/3$ &  $\bar{x}_{Li}+2/3$ &  $z_{Li}$\\
 \end{tabular}
\end{ruledtabular}
\end{table*}

\appendix

\nocite{*}